\documentclass[12pt]{iopart}
\usepackage{iopams,graphicx,color}  
\usepackage[noadjust]{cite}

\begin{document}

\title[Motor effect on filopodia growth]{Modelling the effect of myosin X 
motors on filopodia growth}
\author{K. Wolff$^{1,2}$, C. Barrett-Freeman$^2$\footnote{KW and CBF contributed equally to this work.}, M.~R. Evans$^2$, A.~B. Goryachev$^3$, D. Marenduzzo$^2$}
\address{
$^1$Institut f\"ur Theoretische Physik, Technische Universit\"at Berlin, Hardenbergstra\ss e 36, D-10623 Berlin, Germany\\
$^2$SUPA, School of Physics and Astronomy, University of Edinburgh, Mayfield Road, 
Edinburgh EH9 3JZ\\
$^3$ Centre for Systems Biology, School of Biological Sciences, University of Edinburgh, Mayfield Road, 
Edinburgh EH9 3JR\\}
\ead{katrin.wolff@tu-berlin.de, dmarendu@ph.ed.ac.uk}
\begin{abstract}
We present a numerical simulation study of the dynamics of filopodial growth 
in the presence of active transport by myosin X motors. We employ both a 
microscopic agent-based model, which captures the stochasticity of the growth 
process, and a  continuum mean-field theory which neglects fluctuations. We show
that in the absence of motors, filopodia growth is overestimated by the 
continuum mean-field theory. Thus fluctuations slow down the growth, especially 
when the protrusions are driven by a small number (10 or less) 
of F-actin fibres, {{and when
the force opposing growth (coming from membrane elasticity) is
large enough}}. We also show that, with typical parameter values for
eukaryotic cells, motors are unlikely to provide an actin 
transport mechanism which enhances filopodial size significantly, 
{{unless the G-actin concentration within the filopodium
greatly exceeds that of the cytosol bulk}}.
We explain these observations in terms of order-of-magnitude estimates of
diffusion-induced and advection-induced growth of a bundle of
Brownian ratchets.
\pacs{87.17.Aa,87.16.Ln,87.15.R-}
\end{abstract}

\submitto{\PB}
\maketitle

\section{Introduction}

Cell motility is a fascinating and intricate process~\cite{bray,cellmotilityreview}. 
Largely, cell motion is driven by the dynamics of the actin cytoskeleton, a 
network of semiflexible polymers -- the actin fibres -- interacting with 
molecular motors and with a number of actin-binding proteins~\cite{cellmotilityreview}. 
Actin fibres grow at one of their ends, called the plus or barbed end, and 
shrink at the other, known as the minus or pointed end. At least when cells 
crawl on a 2D substrate, the mechanism through which they move is well 
understood. 
A simplified view is that the growth of actin fibres at the barbed
end pushes the membrane forward, while the contractility due to myosin motors,
which are mainly at the back, ensures that the cell body is dragged 
along~\cite{bray}. In a more detailed description, crawling proceeds via the 
rectification of Brownian fluctuations in the membrane by actin 
polymerisation~\cite{peskin93}. 
{For this to be a viable motility protocol, there has to be sufficient
 ``friction'' with the substrate for  the growing 
fibres to be able to  push without slipping behind.}
In physiological conditions this required friction is provided by focal 
adhesions i.e. protein clusters which attach the cell to the 
substrate~\cite{focaladhesions}. Such structures 
are likely to be absent in 3D, and it has been proposed that contractility 
may have a more primary role in initiating and sustaining three-dimensional 
cell motility, e.g. within a tissue~\cite{rhoda,poincloux,elsen}. 

When a cell crawls on a substrate, it does so by protruding a flat
sheet of material packed with growing actin filaments: this quasi-2D
structure is known as the {\it lamellipodium}~\cite{bray,lamellipodium}.
{{While the lamellipodium is arguably the best documented 
structure in crawling cells, there are a number of other important 
actin-driven protrusions, such as actin ruffles, pseudopodia, podosomes and 
filopodia~\cite{svitkina,mejillano,mogilner,daniels,daniels2,matthew,papoian1,papoian2,papoian3,papoian4,papoian5}. 
We will be in particular concerned with the latter in the current work.}}  
Filopodia are fingerlike protrusions of the cell which are thought to be 
engaged in exploratory cell movements, e.g. to sense the external
environment prior to lamellipodium-associated motion~\cite{bray}.  
They are formed by bundles of actin fibres, which extend and retract 
due to actin polymerising and de-polymerising at the ends of the fibres in 
these bundles. The fibre tips are protected against capping (which would halt 
growth, or extension, of the filopodium) by cytosolic proteins.
The growth of the actin fibres is instead limited by slow transport of 
monomeric actin to the tip of the filopodium, and eventually by membrane 
elasticity which resists the large deformation associated to the formation 
of these protrusions.

{There have been a number of studies on the physics of filopodia 
in recent years~\cite{mogilner,matthew,papoian1,papoian2,papoian3,papoian4,papoian5}.
Mogilner and Rubinstein proposed a mean-field theory to
study filopodia  dynamics in a seminal paper~\cite{mogilner}, and  
Monte Carlo simulations of varying levels of complexity have been 
implemented \cite{papoian2,papoian5} to assess the role of fluctuations 
and of actin-binding proteins. Here we study the growth of filopodia via 
actin polymerisation in two different frameworks. In the first, we assume 
that the growth is driven  solely by the diffusion of monomeric G-actin to 
the tip of the filopodium. In the second scenario, 
following recent work in Refs.~\cite{papoian1,papoian3,papoian4} we explore the possibility that myosin X motors, which are known to be enriched at the tip of  growing filopodia, enhance the transport of monomeric 
G-actin to the tip. One might expect that this second transport scenario 
would lead to  much faster growth of the filopodium, because transport by 
advection should be, at least for large times, much more efficient than unbiased diffusion. 
As was suggested in Ref.~\cite{papoian1} with a microscopic agent-based
model, and as we show here by numerically simulating a continuum
set of equations, this turns out not to be the case.}

In this paper we work on both a microscopic and a coarse-grained level 
by using, respectively, an agent-based model and a set of continuum equations 
of motion. We compare the two frameworks, both qualitatively and 
quantitatively, and identify key  differences in their predictions.
Our main new results are as follows. 

First, we find that 
when the force opposing growth (which comes from membrane elasticity) 
is large (10 pN or more), then the mean-field approach significantly 
overestimates the growth rate of filopodia.  
This is mainly due to
the mean-field theory's  failure to capture correlations between successive 
polymerisation events at large force,
as we show by analysing a simpler
model where the concentration of G-actin is uniform.  
This result is important as it points to potentially significant limitations of mean-field theory when used to describe a
bundle of Brownian ratchets, an approach which is often followed in the literature.

{Second, our results suggest that, given commonly used parameters for actin 
transport and filopodia dynamics, motors are unlikely  to be able 
to speed up 
the growth of such protrusions by simply delivering actin monomers through
advection. 
To reach our conclusions we use a continuum theory which predicts the
growth dynamics of filopodia by coupling the Brownian ratchet dynamics 
to the diffusion and advection of G-actin monomers to the filopodium tip.
{This complements  previous work  on the same topic~\cite{papoian1} which
arrived at similar conclusions using agent-based simulations}.
We also explain these seemingly counterintuitive results on the basis of 
simple order-of-magnitude estimates for the diffusion and advection induced
growth laws.}

{Finally, we critically assess how our results may be changed
if some of our assumptions are modified. For instance we find that,} 
if one were to assume that the diffusion of 
actin monomers in the narrow tube making up a filopodium were significantly
slower than the bulk diffusion (an order of magnitude), then
unbiased diffusion would  become a serious bottleneck for filopodial growth.
{In that case  motor-driven advection could help to  overcome the bottleneck. }
{Potentially, advection could also become relevant if the 
G-actin concentration within the filopodium greatly exceeded that of the cytosol
bulk, suggesting that experiments aimed at determining that concentration
would be very valuable to make further progress theoretically.}

The paper  is structured as follows. In Section 2.1 we introduce the
model, in Section 2.2 we describe the
agent-based microscopic description of the filopodium dynamics, 
whereas in Section 2.3 we outline our continuum model, which is 
based on a set of partial differential equations. 
We then discuss the results obtained by our models in Section 3,
starting from the case where the tip of the actin bundle within the filopodium
grows solely due to the free diffusion of actin from
the bulk of the cell, and then assess the relevance of 
potential actin monomer transport by advection through the action of myosin motors.
Finally, our conclusions are presented in Section 4.

\section{Model and Methods}

\subsection{The model}

The system we consider is a growing filopodium, enclosed within, and
pushing against, a cylindrical cell membrane (radius $r_{\rm cyl}$, see cartoon in Fig. 1). 
The filopodium is a bundle of $N$ actin fibres, each of which is 
assumed to be infinitely stiff~\footnote{As the persistence
length of a single actin fibre is about 17 $\mu$m~\cite{actin}, 
one should note that a force of only a few pN would be enough to buckle a 
single actin fibre of size 100 nm. Nevertheless our approximation of using
infinitely stiff fibres is here justifiable as the persistence length 
of an actin bundle is much larger than that of a single fibre;
it may scale as the number of fibres square if these are appropriately 
crosslinked~\cite{mogilner,matthew}.}
The tip of the filopodium extends via the polymerisation of monomeric, 
or G-actin, into filamentous, or F-actin. Monomeric actin can either reach 
the tip by diffusing within the cytosol, or, potentially, it can be recruited 
there through directed transport by myosin X motors. The motors  are associated with
the filopodium and  we consider them  to move uniformly  along
the filaments at a constant velocity $v$ (we therefore do not directly
model motors). In our framework, G-actin monomers, in addition to diffusion,
 can  attach and detach from the bundle (at rate $k_{\rm a}$ and $k_{\rm d}$ 
respectively). When they are attached, they are transported towards the
tip by myosin X. Finally, the top of the cell membrane diffuses and is 
subject to a load (a force $f$, which {\it in vivo} comes from elastic 
deformation of the membrane and viscous drag). 

\begin{figure}
\begin{center}
\includegraphics[width=12.cm]{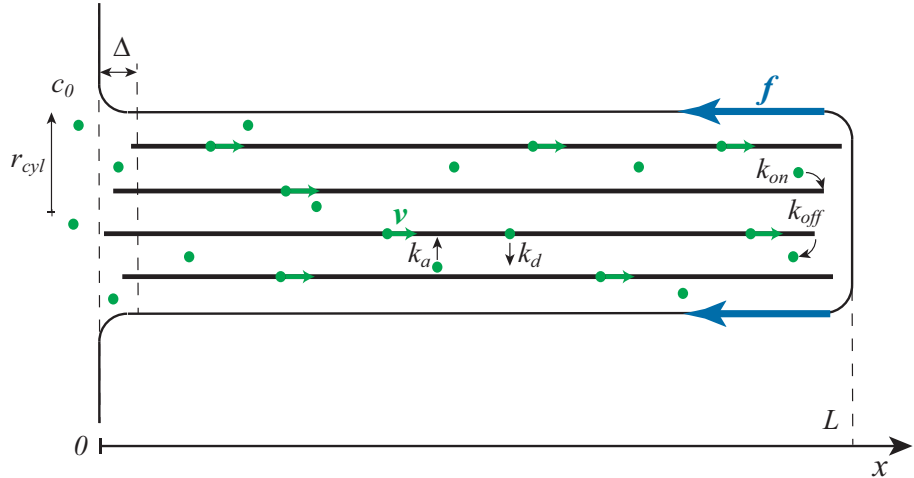}
\end{center}
\caption{Sketch of our model for the growing filopodium inside a cylindrical membrane (whose top, on the right, undergoes Brownian motion against an external load $f$). Symbols are
defined in the text in Sections 2.1 and 2.2.}
\label{cartoon}
\end{figure}

\subsection{Agent-based simulations}

In this Section we describe a microscopic agent-based simulation of {the 
model for  growth of a filopodium  just outlined}.  This approach is useful 
as it can include a relatively high level of detail, and, importantly, it 
incorporates fluctuations in the G-actin density. Our purpose will be to compare
these agent-based simulations to a simplified, and computationally
cheaper, mean-field approach. 

{In the agent-based simulations we explicitly model G-actin monomers diffusing
freely in the 3D space within the filopodial protrusion. This is done by 
attempting, at every time step, to displace monomers randomly by a distance 
chosen uniformly between $-\delta l$ and $\delta l$ along each dimension. 
The resulting diffusion coefficient is equal to $\delta l^2/(6 \delta t)$ 
(see e.g.~\cite{eduardo}), where $\delta t$ is the time step. In the 
simulation, the monomers are taken to be point particles (in other
words we neglect the steric interactions between two G-actin monomers 
or between a G-actin monomer and an F-actin filament).}

The filopodial protrusion is assumed to be cylindrical with constant radius 
$r_\mathrm{cyl}$ and a flat top. Monomers can enter and leave this cylinder 
only via the base. At the base, the G-actin concentration
is held constant at the bulk value $c_0$ ($\sim$ 10 $\mu$M, see 
Table 1)~\footnote{{This boundary condition requires  
inspection of the local concentration at the base, and injection of 
monomers when this concentration falls below $c_0$.}}, while
there are no flux boundary conditions on the lateral 
and top surfaces of the cylindrical filopodium. 

A G-actin monomer that diffuses up to the leading edge of the filopodium can 
polymerise  to become F-actin, if there is a large enough gap between
the fibre tip and the top membrane. In order for 
polymerisation to occur, the distance between the G-actin monomer and the 
tip of one of the F-actin filaments must be smaller than an appropriate 
``polymerisation radius'' (similar concepts arise when simulating
stochastic chemical reactions, see~\cite{ab04}). We also introduce a probability
of polymerisation, with which a G-actin monomer within the polymerisation
range becomes part of the extending F-actin filament. The polymerisation
radius and probability are calibrated so as to give a steady-state 
polymerisation rate, for a fixed G-actin concentration within the filopodium, in
agreement with the experimental rate of 
$k_{\rm on}=10\,\mu\mathrm{M}^{-1}\mathrm{s}^{-1}$. 
This is ensured by choosing a polymerisation radius 
$\delta_{\rm pol}=0.5$ and a polymerisation probability $p_{\rm pol}=0.00917$ 
(both in simulation units).

The top membrane of the protrusion also diffuses. 
To simulate this stochastic membrane motion we postulate that the top
undergoes a random walk against an opposing force $f$
which represents elastic restoring forces.
We use the Metropolis algorithm \cite{ptvf92}, so that we always accept
trial membrane displacements toward  the cell body 
(as long as they  are not impeded by fibres of the filopodium) 
but only accept  membrane displacements away from the cell body 
with probability 
$\exp(-f|\Delta x|/k_\mathrm{B}T)$, where $\Delta x$ is the displacement 
along the positive $x$ direction.
The opposing force  takes a value
$f\sim$ 10-50\,pN,  estimated e.g. in Ref.~\cite{mogilner}.
{Our agent-based simulations essentially follow a kinetic 
Monte-Carlo scheme, which disregards hydrodynamic interactions. Monomers and 
membrane therefore diffuse independently and there is no dragging of monomers 
due to the moving boundary. This is justified by the negligible effects of the 
flow field, $v$, due to the moving membrane compared with the diffusion of
actin monomers (as the Peclet number $v\sigma/D\ll 1$, where $\sigma$ 
and $D$ are the G-actin size and diffusion coefficient respectively).}

{At the growing tip of the filopodium we also model F-actin 
depolymerisation: at a rate $k_{\rm off} \simeq 1$ s$^{-1}$ 
(realistic {\it in vivo}~\cite{carlsson01})
the last actin monomer turns into diffusing G-actin.}

We simulate only the growth of filopodia, not the first emergence of 
protrusions from the cell body, and therefore start filopodia at a finite 
length of $L_\mathrm{start}$. Finally, we model the retrograde flow by 
{retracting}  the fibres by one F-actin monomer every fixed number of time 
steps thereby losing one actin monomer from the base.

As for length and time scales in the simulations, we use the G-actin monomer 
width of $\Delta x$ = 5.4\,nm to set our simulation
length scale~\footnote{Because of the way the actin monomers are stacked in the 
actin fibre, this only grows by $\delta$=2.7\,nm=$0.5\,\Delta x$
upon addition of a G-actin monomer.}. The time scale is set by the 
``Brownian time'' over which G-actin diffuses its own size, 
$\tau=\Delta x^2 / (6D)$, where the diffusion 
constant is $D=5\,\mu\mathrm{m}^2 / \mathrm{s}$. We choose a simulation time 
step of $\Delta t = 0.04\,\tau$. An overview of the model parameters and 
references can be found in Table~\ref{simconsts}.
Note that to simulate membrane diffusion we 
use a maximum step size equal to $0.1$ 
{(in simulation units), corresponding to a diffusion 
coefficient $D_{\mathrm{m}}=1.25$~$\mu$m$^2$~s$^{-1}$. This value
is reasonable for an object of the size of the filopodium diameter, in
an aqueous (rather than an intracellular) environment. In reality,
the effective viscosity of the medium into which the filopodium grows may well 
be larger; however, our results are qualitatively unchanged provided that 
membrane diffusion remains fast with respect to actin polymerisation. This
is true if the dimensionless ratio 
$c(L)\,k_{\rm on}\delta^2/ D_{\mathrm{m}} \ll 1$
at all times,  where $c(L)$ is the G-actin concentration at the tip.
This condition  is
{realistic} for filopodia under physiological conditions. 
Moreover, the condition is  necessary for the ratchet 
equations~\cite{peskin93} to hold.}
\begin{table}[t]
\caption{Model Parameters}
\begin{tabular}{l l l l}
\hline
Symbol & Meaning & Value & Reference\\
\hline
 $r_{\mathrm{cyl}}$ & Radius of filopodium & 100 nm & \cite{sheetz92}\\
$N$ & Number of fibres & $\sim10-30$ & \cite{mogilner}\\
$f$ & Membrane resistance force & $\sim 10 - 50 $pN & \cite{mogilner}\\
$k_\mathrm{B}T$ & Thermal energy & 4.1 pN $\times$ nm & \cite{peskin93}\\
$\delta$ & Actin monomer half-width & 2.7 nm & \cite{peskin93}\\
$D$ & G-actin diffusion constant & $\sim 5\mu$m$^2$/s &\cite{mcgrath98}\\
$D_\mathrm{m}$ & Membrane diffusion constant & $\sim {\cal O}(1)\,\mu$m$^2$/s &\\
$c_0$ & G-actin concentration in & 10 $\mu$M &\cite{mogilner}\\
 & the cell body & & \\
$k_{\rm on}$ & Polymerisation rate & 10  $\mu \textrm{M}^{-1}\textrm{s}^{-1}$ &
\cite{pollard86}\\
$k_{\rm off}$ & De-polymerisation rate at tip & $\simeq$ 1 s$^{-1}$ & \cite{carlsson01}\\
$v_{\mathrm{retr}}$ & Retrograde flow velocity & 10-30 nm s$^{-1}$&
\cite{gardel08}\\
$\eta$ & Geometric conversion coefficient & 18.9 $\mu\textrm{M}^{-1}$
$\mu\textrm{m}^{-1}$& \cite{mogilner}
\end{tabular} 
\label{simconsts}
\end{table}

{Our filopodium is made up of a bundle of static rigid 
fibres, and for simplicity these are imagined to be anchored to the 
underlying cytoskeletal mesh, so that the origin of
the filaments does not move~\footnote{We expect that
allowing fibres to diffuse under the action of a spring which links them to 
their origin (to mimic entanglement or attachment to the cytoskeletal mesh) 
will not change our main results qualitatively. This is supported by
selected simulations within the simplified model of Section 3.2.}.}
{Therefore, in principle, it matters where each fibre is 
initially positioned with respect to the base of the filopodium.}
Placing all the fibres in exact alignment leads to stalling at an 
artificially small value of the force, because all fibres in the bundle 
`lock' at the same distance from the membrane top {and the 
membrane would have to be displaced by an entire monomer half-width to allow
polymerisation}. This is unphysical as in reality the actin filaments can 
displace and bend to accomodate growth at the end~\cite{elasticratchet1,elasticratchet2,comet}.
To overcome this problem, we placed the fibres within a distance $\Delta$ 
of the base. We found that spacing the fibre roots uniformly in
$[0,\Delta]$ and choosing $\Delta=\delta$, the size of the increment of an F-actin filament,  leads to the fastest growth, 
therefore, we have chosen these settings. 
This choice of parameters reproduces the 
stalling force predicted by the mean-field continuum
model which we discuss below. {Positioning the fibres
randomly leads to the same stalling force, and to qualitatively
similar results as the ones reported below, although the growth rate
is quantitatively lower.}

\subsection{Continuum Mean-field equations}
{In this section we introduce a  continuum 
model  to  which  the agent-based simulations may be compared.}
We propose the following set of partial differential 
equations for the dynamics of a growing filopodium:
\begin{eqnarray}
\frac{\partial c_a}{\partial t} & = & 
-v \frac{\partial c_a}{\partial x} + Nk_{\rm a}c_d-k_{\rm d}c_a \label{conca}\\
\frac{\partial c_d}{\partial t} & = &  
D \frac{\partial^2 c_d}{\partial x^2} -Nk_{\rm a}c_d+k_{\rm d}c_a \label{concd}\\
\frac{\partial L}{\partial t} & = & \delta\left[k_{\rm on}e^{-\beta f\delta/N}
c_d(L)+\frac{v c_a(L) \eta}{N} e^{-\beta f\delta/N} -k_{\rm off}\right]-v_{\rm retr}
\label{ratchet}
\end{eqnarray}
where $c_a(x,t)$ is the concentration of advected G-actin, $c_d(x,t)$ 
is the concentration of diffusing G-actin and $L$ is the filopodium length.
{We note that similar equations have appeared in Refs.~\cite{mogilner,papoian2,papoian3}, although the coupling between ratchet dynamics and simultaneous advection and diffusion of G-actin is new to our approach.}

The first equation, Eq.~\ref{conca}, represents the advective transport of G-actin 
and $v$ is the motor-induced advection velocity along the filopodium.
The second and third terms on the right hand side of Eq.~\ref{conca} represent 
attachment of free G-actin to each of the $N$ fibres  with rate $k_{\rm a}$ and 
unbinding of attached G-actin with rate $k_{\rm d}$. The factor $N k_{\rm a}$  
reflects the fact that diffusing actin can attach to a motor on each of the $N$ 
fibres in the bundle.

{Eq.~\ref{concd} represents the diffusive transport of free G-actin, where  $D$ 
is the diffusion coefficient of free G-actin in the cytosol. Again, the second 
and third terms on the right hand side represent attachment of free G-actin and 
unbinding of attached G-actin.
}

{The last of the equations, Eq.~\ref{ratchet}, is the ratchet equation for 
filopodium growth. The terms in the square brackets represent growth and 
shrinkage from processes at the tip whereas the term $-v_{\rm retr}$ represents 
shrinkage due to the filopodium retrograde flow. At the filopodium tip,
$k_{\rm on}$ and $k_{\rm off}$ are the
on and off polymerisation rates   for the
interaction between G-actin and the filopodium actin filaments 
(which are assumed to be uniformly 
covered with myosin X motors) and $vc_a(L) \eta$ is the polymerisation rate
from advected G-actin. For convenience, we write the latter rate
in terms of  $\eta$, a dimensional factor, equal to
$\sim 18.9\, \mu M^{-1} \mu m^{-1}$~\cite{mogilner}, 
which transforms densities per unit volume into
densities per unit length of the filopodium.
The polymerisation rate $k_{\rm on}$  and $vc_a(L) \eta$
are multiplied by the Boltzmann factor
${\rm e}^{-\beta f\delta/N}$, $\beta=\frac{1}{k_BT}$, with $k_B$ the Boltzmann constant and $T$ the temperature,
and $f\delta/N$ is the energy cost of extension against the constant opposing force $f$ of the resisting membrane.
We take $\delta$, the size of the increment of an F-actin filament, to be equal 
to 2.7 nm, half the size of a G-actin monomer.
Note that in the ratchet equation for $L(t)$, Eq.~\ref{ratchet},
the increase in the length of the filopodium comes from two separate additive 
terms that correspond to the diffusing and advected populations. While the 
first term which comes from diffusion is standard~\cite{mogilner}, 
the second, advection contribution is new---note that the 
increase due to advection is inversely proportional to the number of 
fibres, as we need $N$ advected monomers to increase the whole filopodial 
length by the size of an actin monomer. {The advection
contribution follows from assuming that monomers extend the filopodium
as soon as they are advected to the tip. In principle {one might include  a reaction rate for this process,
but this would not qualitatively affect our results.}}
{We also highlight at this point the
mean field ``load sharing'' approximation in Eq.~\ref{ratchet},
according to which the growth of a bundle can be written by 
mapping $k_{\rm on, off}\to N k_{\rm on,off}$, $\delta \to 
\delta/N$ in the equation valid for a single 
fibre~\cite{mogilner,daniels,matthew}. As we shall see, this approximation 
may lead to discrepancies between the continuum theory and the agent based 
simulations. We will further discuss this fact in Section 3.2.}

We consider the following boundary conditions:
\begin{eqnarray}\label{bc1}
c_a(x=0,t)+c_d(x=0,t) & = & c_0 \\
\label{bc2}
\frac{c_a(x=0,t)}{c_d(x=0,t)} & = & \frac{Nk_a}{k_d} \\
\label{bc3}
-\left[\frac{D \partial c_d(x,t)}{\partial x}\right]_{x=L(t)} & = & 
\frac{N}{\eta}\left[k_{\rm on}c_d(L)e^{-\beta f\delta/N}-k_{\rm off}\right],
\end{eqnarray}
where $c_0$ is the bulk concentration of G-actin, 10 $\mu$M.

{At $x=0$, i.e. at the base of the 
filopodium,  boundary condition  (\ref{bc2}) states that the advected  and freely diffusing populations are in 
equilibrium and 
boundary condition  (\ref{bc1}) states that  the total density is the typical bulk 
concentration of G-actin. This choice of boundary conditions  has   previously been made
in the literature in Ref.~\cite{papoian1}, where it was argued that
this boundary condition is most consistent with existing
experimental data. An alternative choice of boundary condition would be to fix $c_d(0)$, rather than 
$c_d(0)+c_a(0)$,  to $c_0$, and to take  $c_a(0)$  to be 
proportional to $N$, as more filaments provide more binding sites. However, this would 
also require the G-actin density in filopodia to be several-fold larger than in 
the bulk, a fact which has not been reported to date to our knowledge. This 
alternative choice of boundary condition \textit{would} modify our conclusions and we will 
discuss it again in Section 3.4 when we comment on the implications of our 
results for the growth of filopodia in presence of myosin X motors, but we
believe it to be less realistic than the boundary conditions in Eqs.~\ref{bc1}
and \ref{bc2}.}

On the other hand, the boundary condition Eq.~\ref{bc3} at $L(t)$, i.e. at the 
tip of the filopodium, states that there is a sink for the diffusing G-actin,
due to polymerisation. {This sink term is formally the same as the purely 
diffusive term proposed in the previous work~\cite{mogilner,papoian1},
which comes from our assumption (discussed above) that the exit flux of 
advected actin at the tip of the filopodium is 
$v e^{-\beta f\delta/N} c_a(L)$ (hence drops out of the equation for the
diffusing G-actin sink).}

Note that we do not include  any exclusion interaction between
the motors. The exclusion interaction has revealed interesting properties
in a number of molecular motor systems on dynamic filaments
e.g. fungal hyphal growth \cite{SEPR}, elongating actin filaments \cite{NFC} and
extraction of membrane tubes by motors \cite{TEK}.  In principle, we could 
include exclusion  by turning the advection equation into a Burgers equation 
with a reaction term -- as was incorporated in  recent work on 
filopodia ~\cite{papoian3} for the density profiles of 
motors on actin filaments. Our choice to not model exclusion is justified in our 
context as its effects would only be important if the concentration of bound 
monomer were as high as 1 per filament per 5 nm (corresponding to 50\% of the 
filopodial length being associated with advected G-actin). This requires, for a 
filopodium made up of 10 fibres, a 200 micromolar concentration of G-actin, 
which is far above the typical bulk G-actin concentration (around 10 micromolar) 
considered in our calculations.

Finally, we note that we can estimate the maximal length 
of the filopodium from the above equations,
under the assumption that there is no advection (we follow
exactly the same procedure  used in~\cite{papoian2}, repeating
the intermediate steps for the reader's convenience).  We assume
that the density profile in the steady state will be a linearly decreasing
function of $x$, and the gradient can therefore be reasonably approximated by:
\begin{equation}
 \frac{\partial c}{\partial x} \simeq \frac{c(L)-c_0}{L}.
\end{equation}
(This assumption is backed up by observations of density profiles
in numerical simulations of Eqs.~\ref{conca} and \ref{ratchet}.)
We then use the boundary condition at the tip to derive an
expression for the actin concentration there, $c(L)$:
\begin{equation}
 c(L)=\frac{LNk_{\rm off}+c_0D\eta}{D\eta+LNk_{\rm on}\exp(-f\delta/Nk_BT)}.
\end{equation}
Substituting this result into the equation for $L(t)$ and requiring
$\textrm{d}L/\textrm{d}t=0$ yields  the steady state, or maximal, length:
\begin{equation}\label{maxlength}
 L_{\rm{max}}=\frac{D\eta}{Nk_{\rm{on}}}\left[\frac{\delta k_{\rm on}
c_0}{v_{\rm{retr}}}-\left(\frac{\delta
k_{\rm off}}{v_{\rm{retr}}}+1\right)\textrm{e}^{\frac{f\delta}{Nk_BT}}\right].
\end{equation}

Using the values in Table 1, we find the maximal length to 
be $\sim 1-10 \mu$m.  

\section{Results}

\subsection{Filopodia growth in the absence of motors: comparison between
agent-based and continuum models}

In this Section we compare the results obtained from numerical integration of 
Eqs.~\ref{conca}-\ref{ratchet} (which we refer to as the
continuum mean-field theory), with those from agent-based 
simulations in the absence of motors. This will help to understand whether the 
mean-field model is a good approximation for the system it attempts to describe 
and, if so, what ranges of parameters it works well for. 
To this end, we compare the filopodium growth (lengths as a function of time) 
obtained with the two methods. 
In general, as we will show below, the agreement is qualitatively good, 
at least for physiological values of the parameters. There are however 
significant quantitative differences which we will also address.  

As should be expected, the agreement between agent-based simulation and the 
continuum mean-field theory 
is best for a large number of fibres $N$ (see Fig.~\ref{figure1}). This is 
due to the effect of fluctuations, neglected in the mean-field 
continuum approach,  which are {more significant} for  a small number of fibres. 
Both  the mean-field theory and the agent-based simulations
predict  that the filopodium grows faster for an intermediate number of fibres
(see Fig.~\ref{figure1} where the bundle of $N=10$ fibres 
grows faster than that with $N=1$ or $N=30$). 
This non-monotonic behaviour can be explained intuitively  by noting that
for a large number of filaments the bundle needs more actin monomers to 
fuel its growth. On the other hand, growth is considerably reduced in the 
case of very few fibres as the bundle is more sensitive to the 
action of the external force: this effect enters through  the factor of $f/N$ in 
the Boltzmann factor in the ratchet equation in Eq.~\ref{ratchet}
(see also~\cite{mogilner,daniels,matthew}). 

\begin{figure}
\begin{center}
\includegraphics[width=12.cm]{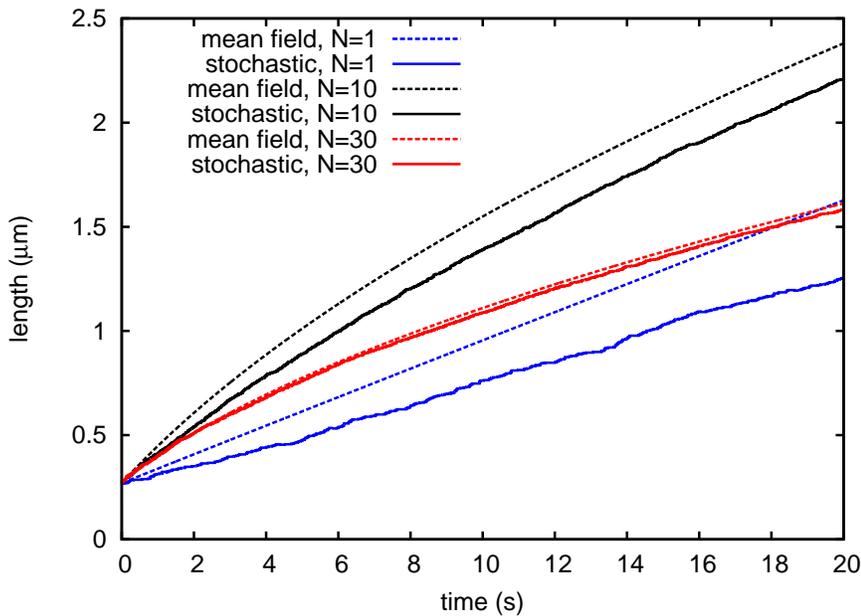}
\end{center}
\caption{Comparison of agent-based (solid lines) and mean-field
(dashed lines) simulations, in the absence of advection by
molecular motors. The only transport channel for G-actin is
through unbiased diffusion. Membrane resistance force is $f=2$\,pN,
concentration $c_0=10\,\mu$M, depolymerisation rate $k_{\rm off}=1$\,s$^{-1}$.
The number of the filaments in the bundle is varied (see legend).
The mean-field approximation improves for larger values of $N$, 
for which stochastic fluctuations are less important. The data
also show that there should be an optimum number of filaments
at fixed monomer concentration -- 10 fibres grow faster than either
30 or a single filament. This result is fully consistent with the findings
in Ref.~\cite{mogilner}, which were obtained with a continuum mean-field
theory.}
\label{figure1}
\end{figure}

Next, we investigate the effect of varying the membrane force $f$ at a fixed
number of fibres $N=30$ (a reasonable assumption for typical
filopodia {\it in vivo}~\cite{svitkina,mogilner}). 
For small values, $f=1-5$~pN, the  hindering of growth  is rather minimal,
in both the mean-field and the agent-based model. 
When $f=5$~pN, the agent-based simulation begins to be affected and 
growth is reduced. We find that the agent-based model is affected   by increasing force 
more strongly than the mean-field model, leading to a marked
discrepancy between the two approaches (see Fig.~\ref{figure2}). 

Increasing the force further decreases the growth rate until the stalling force 
is reached. The prediction for the stalling force with $N=30$ is given by 
Eq.~\ref{ratchet} as $f_s=210$~pN and both the continuum model solution and the
microscopic dynamics agree on this.
{The greater sensitivity of the growth rate on the applied force in the
agent-based simulation is due to the mean-field approximation 
implicit in the ratchet equation (\ref{ratchet})
i.e. in going from a ratchet equation for a single filament to an equation for a bundle of $N$ filaments
we have simply replaced   $\delta\to\delta/N$, $k_{\rm on}\to N k_{\rm on}$ and
$k_{\rm off}\to N k_{\rm off}$ 
 (see also Refs.~\cite{mogilner,daniels,matthew}).
This point will be discussed further in Section~\ref{sec:breakdown}. }

\begin{figure}
\begin{center}
\includegraphics[width=12.cm]{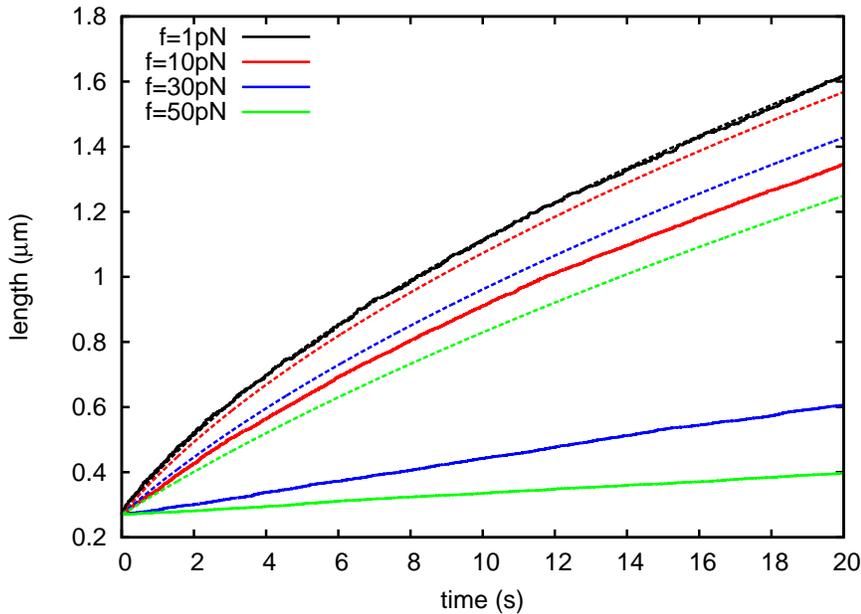}
\end{center}
\caption{Comparison of agent-based (solid lines) and mean-field
(dashed lines) simulations, in the absence of advection by
myosin motors. The number of filaments is fixed at $N=$30, while
the concentration of G-actin in the bulk (at the base of the
filopodium) is $c_0=$10 $\mu$M, and the depolymerisation rate is
$k_{\rm off}=$1 $s^{-1}$. The membrane resistance force is varied.
The mean-field approximation gets worse for increasing force, 
where it severely underestimates the slowing down induced by the
external force.}
\label{figure2}
\end{figure}

Increasing bulk concentration $c_0$ speeds up the growth (data not shown), as 
would be expected, but has little effect on the agreement between the mean-field
theory and the microscopic simulation. 

Up to now we have neglected the retrograde flow, $v_\mathrm{retr}$
(see Section 2.1 and 2.2). This is not realistic as in practice 
actin polymerisation
is always accompanied by a retrograde flow of the network~\cite{bc02}.
If we now introduce a non-zero  $v_\mathrm{retr}$, we find that the 
system can reach a steady state as expected from the analysis of the continuum 
equation without the advection proposed above (and paralleling that in 
Ref.~\cite{mogilner}). For small values of the retrograde flow,
$\sim$10 nm s$^{-1}$ (typically quoted in 
experiments, see e.g.~\cite{bc02}), the 
continuum model predicts that the system should take about 1000\,s to reach 
a steady state. For greater values, such as $v_\mathrm{retr}=70$ nm 
s$^{-1}$, employed in \cite{mogilner,papoian2} to describe filopodia
emerging from a lamellipodium, {the steady state would be attained an order of 
magnitude sooner}. 
In Fig.~\ref{sslengths}, we compare the approximate steady-state 
solution found in Section 2.2, see Eq.~\ref{maxlength}, 
with the steady-state lengths found by direct 
numerical simulation of the mean-field equations --
the estimate and the exact value are in excellent agreement.
{We also simulated growth in the agent-based model
with retrograde flow (open symbols in Fig.~\ref{sslengths}).
The results reinforce our main previous finding, that for larger forces the 
continuum model overestimates growth rates (Fig. S1) and also, in this case, 
steady-state lengths.}

\begin{figure}[t]
\vspace{0.75cm}
 \centering
 \includegraphics[width=12.cm]{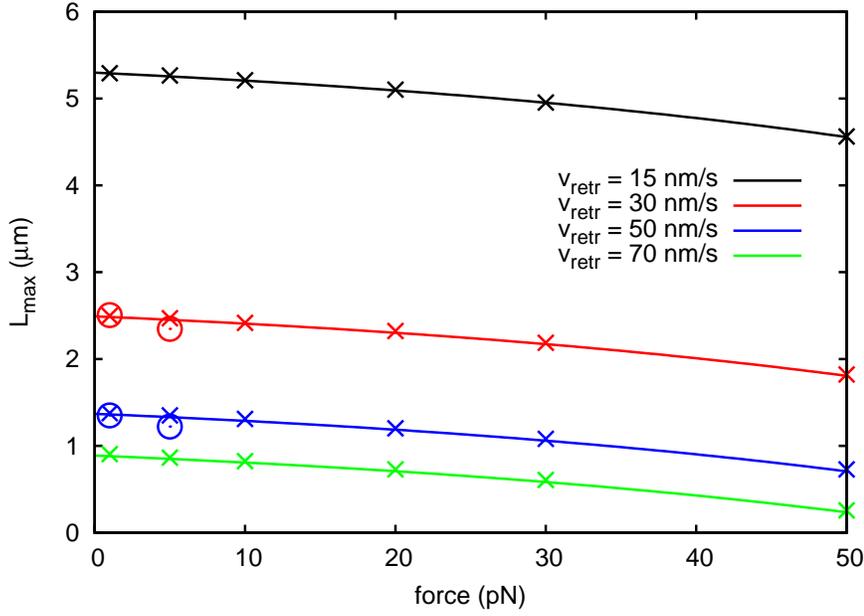}
 \caption{Steady-state filopodium length as a function of membrane resistance
force from analytic prediction (solid lines), numerical integration of the 
mean-field continuum equations (crosses) {and agent-based 
simulations (open circles)} for various values of the membrane
resistance, $f=1.0, 5.0, 10.0, 20.0,30.0,50.0$~pN. The retrograde flow rate is
also varied $v_{\rm retr}=15,30,50,70$~nm s$^{-1}$. Other simulation parameters 
are $N=30$, $k_{\rm off}=1$~s$^{-1}$ and $c_0=10$~$\mu$M.}
 \label{sslengths}
\end{figure}

In this section we have seen that the agreement between the agent-based 
model and numerical simulation of the mean-field model is reasonably close for 
realistic parameter values. 
For small numbers of actin filaments the discrepancy between the models 
grows (as expected) but qualitatively the predictions still agree. 
The main parameter 
which appears to affect the system differently in the simulation and continuum 
theory is the load force $f$ at the filopodium tip. The agent-based model is 
affected by an increased force much more than the 
continuum mean-field theory, and filopodium 
growth at large force is over-estimated in the latter theory. 
{Although it is natural to attribute the  discrepancy to   neglect of
fluctuations in the mean-field theory, it is not directly obvious how this actually occurs.
We will investigate it 
further in the next subsection.}

\subsection{Breakdown of mean-field theory at large load}\label{sec:breakdown}

{We now  examine  in more detail  the reasons} for the 
quantitative inaccuracies of the mean-field approach. The 
discrepancies at low values of $N$ can be understood as
being due to the fluctuations in the actin monomer concentration at the tip, 
which are incorporated in the agent-based simulations but 
are absent in the continuum equations. However, this cannot be the 
reason for the breakdown we observed at large load (Fig. 3), as 
fluctuations should not depend dramatically on the force applied to the 
membrane and limiting filopodium growth.

To explore this issue further, we have performed simulations in which
the dynamics of monomeric actin is not explicitly include, instead it is always taken to be  at the  bulk concentration, $c_0$. Therefore, the growth law 
of the filopodium length, $L(t)$, in
the mean-field approximation used above
obeys a simpler equation, namely
\begin{equation}
\frac{\partial L}{\partial t} = \delta\left[k_{\rm on}c_0 e^{-\beta f\delta/N}
-k_{\rm off}\right]-v_{\rm retr}.
\label{ratchetconstantactin}
\end{equation}

The agent-based simulation is run with $N=30$ actin filaments, each of which
can polymerise at rate $k_{\rm on}c_0$ and depolymerise at rate
$k_{\rm off}=k_{\rm on}c_0/100$. We disregard retrograde flow, and
initialise the filaments with a displacement along their direction
equal to $\delta/N$ between each other -- this is 
to avoid the locking problem mentioned in Section 2.1.
The comparison between the simulation results and 
Eq.~\ref{ratchetconstantactin} in Fig.~\ref{figure3} clearly shows 
that the discrepancy is
already there at the level of a simulation which disregards any
fluctuation in the local monomeric actin concentration. 
{The breakdown is instead due to the approximation that
having a bundle of $N$ fibres, as opposed to a single fibre, 
can be simply taken into account by setting
$k_{\rm on}\to k_{\rm on}N$ in the single fibre equation, as this assumes that all fibres have the
same chance of polymerising.} While not a bad assumption for
small $f$, this approximation breaks down dramatically for large $f$, 
as the filaments essentially grow one at a time. This is because the
fibre farthest from the membrane is by far the most likely to 
elongate by exploiting a gap which appears due to membrane diffusion.
Consequently, the polymerisation events in the dynamics are highly correlated and the mean-field approximation breaks down. 
\begin{figure}
\begin{center}
\includegraphics[width=12.cm]{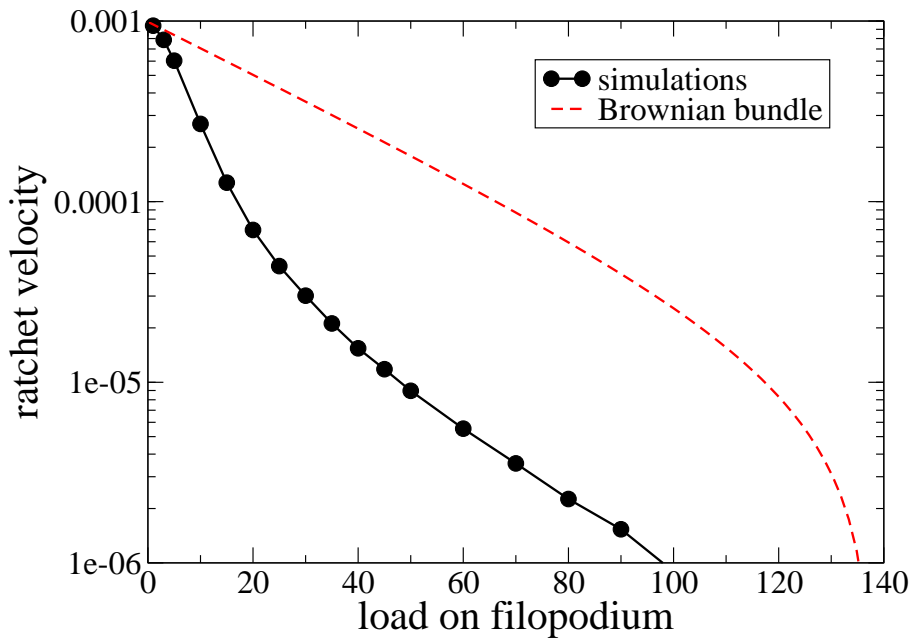}
\end{center}
\caption{Comparison between the  mean-field prediction (dashed lines) and simulations (points)
for the velocity of a bundle of Brownian ratchet,
for $v_{\rm retr}=0$ and $k_{\rm on}c_0/k_{\rm off}=100$. It can be
seen that the mean field significantly overestimates the numerical data.
Data are shown in simulation units: one force unit corresponds to
about 1.5 pN.}\label{figure3}
\end{figure}

\subsection{Filopodia growth in the presence of motors: parameter choices}

From the numerical results in Section 3.1 and Fig. 4, we can see 
that for the realistic values of parameters the maximal length that can be 
supported by this diffusion-limited process is quite small, even when
neglecting the effect of fluctuations which, as we showed in Section 3.2, 
further limit the steady-state length of the filopodium. There must 
therefore be another mechanism that allows the filopodia to reach lengths as 
large as 40~$\mu$m~\cite{mogilner}.
One possibility, that we explore in this section,  is that larger lengths  are made possible via the directed transport of actin monomers by myosin motors~\cite{papoian2,papoian3}.  
which have been associated with filopodia growth for a long 
time~\cite{bc02,bpc01, wwcmj96}.  Whereas a long bundle will have to wait a
significant amount of time for an actin monomer to reach the tip by diffusion
alone, it might be possible that the advective transport process will be 
much faster.  

First  we discuss  the choice of parameters that we make
to study filopodial growth with
motors.
In particular, we have considerable freedom in choosing the attachment and 
detachment rates ($k_a$ and $k_d$ respectively). 
In what follows, we for simplicity ensure  (unless specified otherwise) 
that $Nk_a=k_d$ so that $c_d\sim c_a$ at the filopodial base, 
which is reasonable given the recent 
experimental data~\cite{nrncbsr08}.  
Note that in~\cite{papoian2}, the authors consider a very large
range of $k_d=1-3000$~s$^{-1}$, but as we will see below, we require much 
smaller values of $k_d$ in order to allow motors to aid filopodial growth.  
The remaining key parameter in the advection simulation is the
motor advection, $v$. Myosin X is known to be a highly processive motor,
however the details of its interactions with G-actin are poorly 
understood~\cite{papoian2}. Here we will present results with a value
of $v$, in the $\mu$m/s range, which is at the high end of the
biologically relevant range~\cite{papoian2}.

In the following section, we analyse the numerical results obtained
by considering myosin-aided advection in our mean-field theory in
Eqs.~\ref{conca}-\ref{ratchet}.

\subsection{Filopodia growth in the presence of motors}

We now discuss the filopodium dynamics predicted by Eqs.~\ref{conca}
and \ref{ratchet}, when motor-induced transport is included. 

In analogy with the presentation of the results in Section 3.1, we
first consider the (physically unrealistic) case in which there is 
{\it no} retrograde flow (parameters not discussed below are set 
according  to Table 1).
Fig.~\ref{no-retrograde-length} compares growth dynamics of a 
filopodium (made of $N=10$ fibres) in the absence and presence of 
myosin-mediated transport; {while Fig. S2 shows 
some typical corresponding density 
profiles of diffusing and advected monomers.}
When motors are included the initial growth rate of the filopodium is actually
slower. Although this result may at first sight seem surprising, it is 
consistent with the findings of Ref.~\cite{papoian1}, which studied a similar system with a spatial Gillespie algorithm. 
The reason for such a behaviour can be appreciated by comparing the initial 
growth rate coming from diffusion and motor advection respectively. 
The former can be estimated as (parameters as in 
Fig.~\ref{no-retrograde-length})
\begin{equation}
\left[k_{\rm on} c_0 e^{-\beta f\delta/N} -k_{\rm off}\right] \delta
\sim 137 \, {\rm nm \, s}^{-1}.
\end{equation}
The latter, the initial growth rate when considering also advection, can be 
estimated as
\begin{equation}\label{condition1}
\left[k_{\rm on} c_d(0) e^{-\beta f\delta/N}+\frac{v c_a(0)
e^{-\beta f\delta/N}\eta}{N} -k_{\rm off}\right] \delta \sim 74 \, {\rm nm \, s}^{-1}.
\end{equation}
Therefore, paradoxically, sequestration of G-actin monomers by myosin X motors 
advecting along the actin bundle initially {\it slows down} rather than 
accelerates the length growth. 

By using this argument, it is straightforward to find a criterion  on
the parameters  for  the filopodium  to initially grow
faster with advection. 
For this to be the case, one requires:
\begin{equation}\label{condition2}
 k_{\rm on} c_d(0)+\frac{vc_a(0)\eta}{N}>k_{\rm on} c_0.
\end{equation}
Using the boundary conditions Eqs.~\ref{bc1}--\ref{bc3}, and re-arranging the 
terms, we can find that this condition is linked to the value of the
 following dimensionless number:
\begin{equation}\label{dimless1}
 \lambda=\frac{v\eta}{Nk_{\rm on}}.
\end{equation}
If $\lambda>1$, the initial growth will be greater with motors.
Using values in Table 1, we find  that $\lambda=1$ requires
$v\sim 5$ $\mu$m s$^{-1}$, which is unlikely for myosin X advection. 
We should stress that, while Eq.~\ref{condition1} holds in general,
Eq.~\ref{condition2} exploits the assumption that $c_a(0)+c_d(0)=c_0$,
as used previously in  in Ref.~\cite{papoian1}.
{A different boundary condition at the base of the filopodium 
may therefore affect our conclusion. In particular using $c_d(0)=c_0$ and
taking $c_a(0)$ proportional to the number of filaments $N$ leads to an 
enhancement of the growth rate by motor advection (see Fig. S3, where the
same value of $v$ used in Fig.~\ref{no-retrograde-length} is used). However,  as 
mentioned when introducing our continuum model, this assumption would 
 require the G-actin concentration to be several-fold larger 
than in the bulk, a hypothesis for which there is no clear evidence to date~\footnote{We note here that even if $c_a(0)\eta$ 
approaches the jamming density on the bundle (equal to about one 
transported G-actin monomer every 5 nm in each filament), advection 
leads to an initial greater flux with respect to diffusion for 
an advection velocity of $v>0.5$ $\mu$m s$^{-1}$ (as in 
Fig.~\ref{no-retrograde-length} and Fig. S3). This is still a 
very large value if we consider the fact that
the association between G-actin and myosin X motors may be
reversible~\cite{papoian2}.}.}

In the absence of retrograde flow, as noted also in Section 3.1, there can be 
no steady state. Under this condition advection eventually leads to longer 
filopodia, as it yields linear growth if $c_a(L)\ne 0$, 
as opposed to $L(t)\sim t^{1/2}$ when diffusion is the only transport
mechanism. However, as is apparent from Fig.~\ref{no-retrograde-length}, the
crossover between diffusion-dominated and advection-dominated growth 
occurs for unrealistically high values of the filopodial length 
(around 20 $\mu$m in Fig.~\ref{no-retrograde-length}, even though a rather large 
advection velocity, $v=0.5~\mu$m~s$^{-1}$, is assumed). These lengths are likely 
to be irrelevant for filopodia {\it in vivo}, as elasticity would halt the 
growth much sooner. 

\begin{figure}[t]
\vspace{1cm}
 \centering
 \includegraphics[width=12.cm]{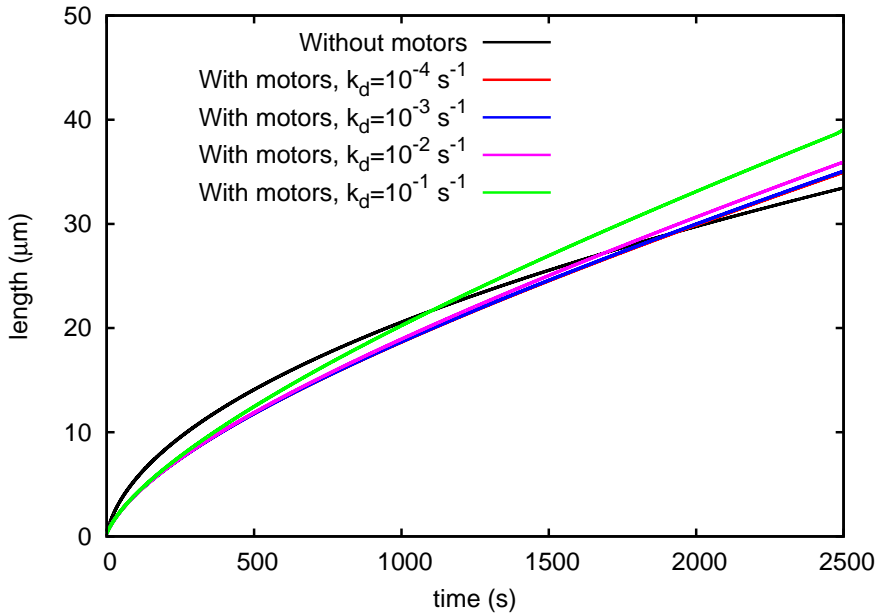}
 \caption{Filopodium length as a function of time from numerical integration of
Eqs.~\ref{conca}--\ref{ratchet}.  
Initially, adding myosin motors slows down filopodial
growth, see discussion in the text.
Simulation parameters are
$N=10$, $f~=~10$~pN, $k_{\rm off}=1$~s$^{-1}$, $k_{\rm on}=10\mu$M$^{-1}$s$^{-1}$, $a_0=10$~$\mu$M and advection velocity $v=0.5~\mu$m~s$^{-1}$}
 \label{no-retrograde-length}
\end{figure}


Also when $v_{\rm retr}$ is non-zero, and a steady state can be reached, 
the size and growth rate of filopodia are not significantly enhanced
by motor transport for $v=0.5$ $\mu$m s$^{-1}$ 
(see Fig.~\ref{retrograde-length}). 
For small values of $\lambda$ and of the kinetic 
 constants $k_a$ and $k_d$
 filopodia grow longer with diffusional transport alone, even at late times. 
On the other hand, advection   eventually leads to  longer bundles when $k_d=Nk_a$ is 
larger (e.g. 0.1 s$^{-1}$, see  Fig.~\ref{retrograde-length}).
Increasing $\lambda$ does lead to a dramatic difference in the kinetics,
however, as previously mentioned, this would imply  that motors
move at unrealistically high speed.


\if{To get a feel for whether or not advection can lead to an advantage over
diffusion, for filopodial length in steady state (hence
at {\it late} rather than early times), we propose another
dimensionless control parameter, 
\begin{equation}
 \theta=\sqrt{\frac{D}{Nk_a}}\left(\frac {k_d}{v_r}\right).
\end{equation}
Here, $v_r$ is the ratchet velocity [equal to $(\delta\frac{v\eta c_a(0)}{N} e^{-\beta f\delta/N}-v_{\textrm{retr}})$], which measures the ratio of the
average length diffused by G-actin in 3D, $\sqrt{D/(Nk_a)}$, over the average 
advection length a monomer spends on the filopodium between an attachment
and a detachment event, $v_r/k_d$.}\fi


\begin{figure}[t]
\vspace{0.75cm}
 \centering
\includegraphics[width=12.cm]{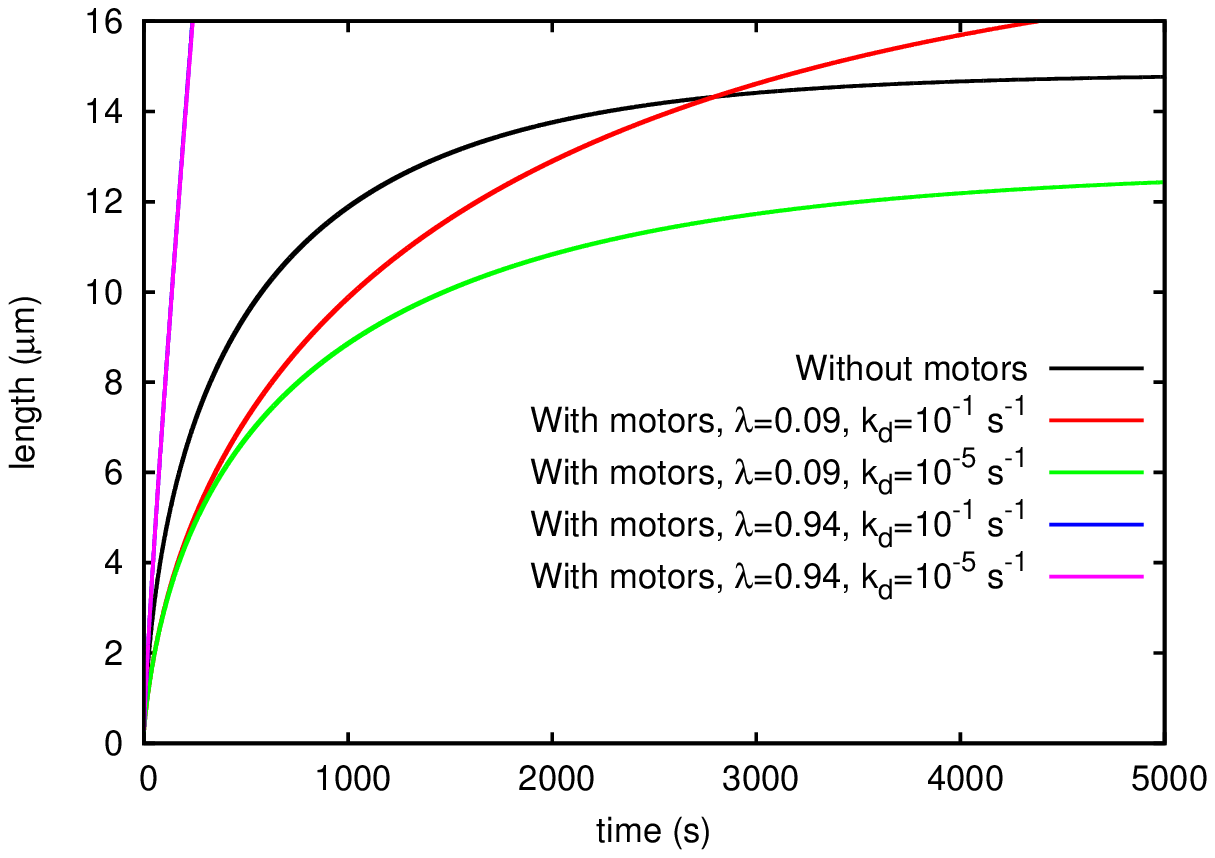}
 \caption{Filopodium length as a function of time from numerical integration of
Eqs.~\ref{conca}--\ref{ratchet}.  
With $v_{{\rm retr}}=15$~nm~s$^{-1}$, the system can
reach a steady state.  In general, including myosin transport results in a
shorter steady-state length.  We show two extreme values of $k_d=10^{-1},
10^{-5}$~s$^{-1}$ for $\lambda=0.09, 0.94$ in green, red and yellow, blue
respectively.  If $\lambda$ is sufficiently large, $k_d$ has little effect
at least in the physilogical range of filopodial length.  If $\lambda$ is
small, the strength of attachment/detachment dictates whether or not
steady state is reached at all.  Simulation parameters are $N=10$, $f=10$~pN,
$k_{\rm off}=1$~s$^{-1}$, $a_0=10$~$\mu$M, $v=0.5$ , $5$
$\mu$m~s$^{-1}$}
 \label{retrograde-length}
\end{figure}

Figs.~\ref{no-retrograde-length} and ~\ref{retrograde-length} strongly
suggest that, with a realistic choice of parameters (see Table 1), 
G-actin advection by myosin X hinders, rather than enhances, filopodial
growth. 
It is, however, possible that the diffusion constant 
$D=5$~$\mu$m$^2$~s$^{-1}$ has been over-estimated in
the literature, where {\it in vitro} experiments might not account for the 
level of macromolecular crowding occurring 
{\it in vivo}~\cite{e01}.  If this were
the case, advection might become more relevant.  In 
Fig.~\ref{reduceddiffusion}, we see
that with $D=0.5$~$\mu$m$^2$~s$^{-1}$ and a small detachment rate
$k_d=10^{-4}$~s$^{-1}$, the system with advection takes over before
$L\sim3$~$\mu$m.  
This represents an almost ten-fold decrease in the crossover point
from the previous case when $D=5$~$\mu$m$^2$~s$^{-1}$, and shows that given the
right parameters, it might be possible for the myosin motors to have a 
positive effect on filopodial growth. \if{The amount of tuning which is
needed to position the cell in this region of parameter space is 
however so significant that we would deem such a scenario rather unlikely.}\fi

\begin{figure}
\vspace{0.75cm}
 \centering
\includegraphics[width=12.cm]{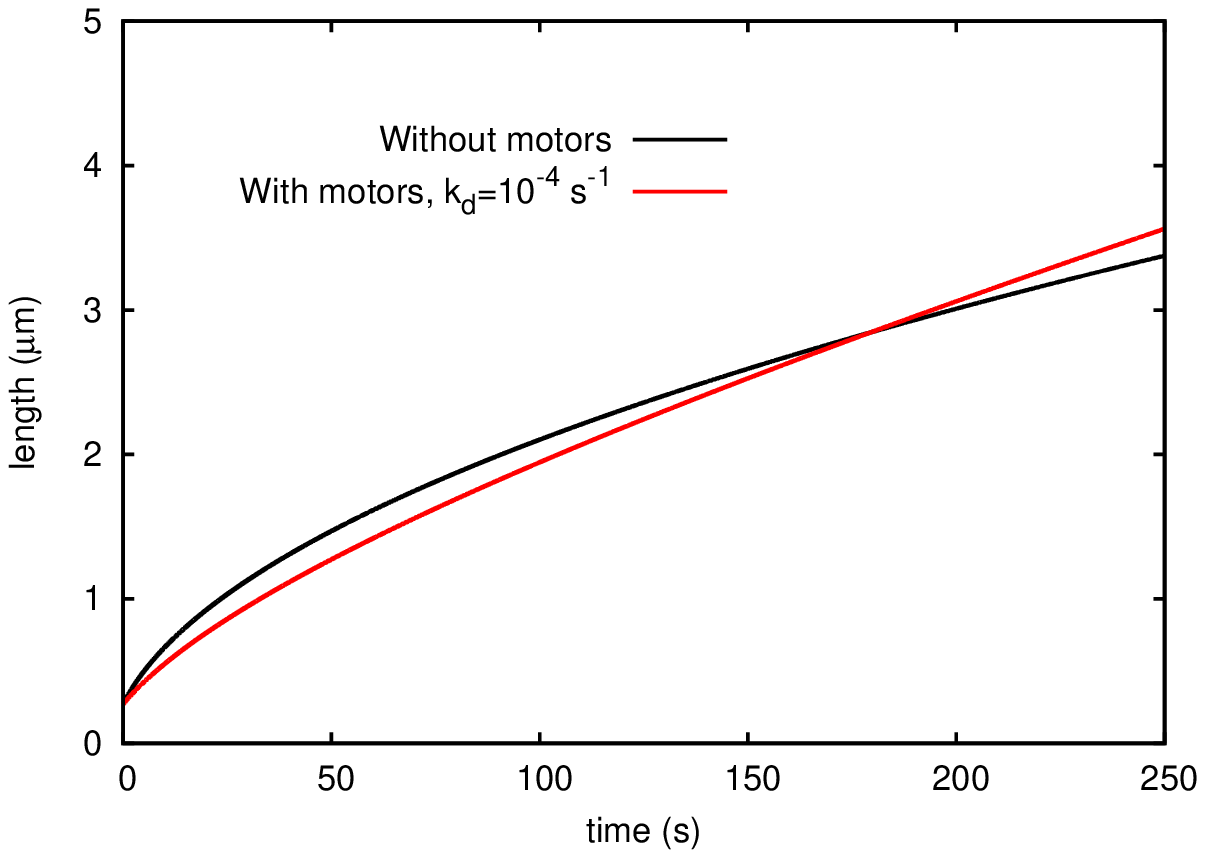}
 \caption{Filopodium length as a function of time from numerical integration of
Eqs.~\ref{conca}--\ref{ratchet}.  Reducing the diffusion constant to
$D=0.5$ $\mu$m$^2$~s$^{-1}$ can make the myosin motors much more relevant as the
crossover point is now below $L\sim3$ $\mu$m.  Simulation parameters are $N=10$,
$f=10$~pN, $k_{\rm off}=1$~s$^{-1}$, $a_0=10$~$\mu$M, $v=0.5$
$\mu$m~s$^{-1}$.}
 \label{reduceddiffusion}
\end{figure}

\section{Conclusions}

In conclusion, we have presented a systematic study
of the growth of filopodia both in the absence and in
the presence of active transport of actin monomers by
myosin motors. We have compared the predictions of 
a set of continuum partial differential equations based on
a mean-field approximation to direct simulations of an agent-based
description resolving the position of each of the fibres, of the top
membrane limiting the filopodium, and of monomeric actin diffusion/advection
within the filopodium.

Our main results may be summarised  as follows. 

First, we found that the mean-field theory and the agent-based simulations are 
in general, qualitative  agreement, but we highlighted
some significant quantitative discrepancies, either for small 
bundles or for large forces acting on the filopodium tip. The former observation is expected, and is
simply due to the neglect of fluctuations in the mean-field theory;
{{a similar  effect has also been noted in Ref.~\cite{papoian2}.}
The latter discrepancy has a different origin,
and we have shown that it appears even in a model
considering a well-stirred environment in which depletion of
G-actin monomers is not taken into account (Section 3.2). We have 
{demonstrated} that the mean-field theory for bundles under a high opposing
force breaks down. This is because the continuum model does not account
for the fact that in the microscopic
description fibres elongate one at a time i.e.
essentially it is only  the fibre end farthest from the membrane which  elongates when
a sufficiently large gap appears. To the best of our knowledge,
this shortcoming of the mean-field theory had not previously been
identified, in spite of the widespread use of the 
ratchet equation Eq.~\ref{ratchet} to describe the growth 
of filopodia~\cite{mogilner,daniels,matthew}.

Second, we found that, surprisingly, with parameter values taken from
the recent literature, myosin-directed transport of actin monomers 
does not effect an increase in the growth rate and steady state
length of filopodia. This can be rationalised quite simply on
the basis of some order-of-magnitude estimates, from which it
appears that, given the accepted values of actin diffusion, 
polymerisation rate and myosin X velocity, the rate of advection-driven
growth is notably smaller than that of diffusion-driven growth
(unless the filopodium is unrealistically long).
However, if we assume that crowding and confinement within the filopodial tip
lead to a smaller diffusion coefficient for actin monomers, then 
motors {\it could} play a role, and lead to more efficient growth
of filopodial protrusions. Furthermore, advection
would also become much more relevant if it turned out that 
the total density of G-actin, including both diffusing and bound
monomers, were much larger than the bulk intracellular concentration
of actin monomers. It would be interesting to probe this possibility
experimentally in the future.

We stress that our results in no way imply that myosin
X transport is in general irrelevant for the physics of filopodia.
On the contrary, it may well be that motor-driven transport is necessary
for molecules, such as VASP, which are involved in the maintenance 
of the filopodial bundle~\cite{mejillano}; we have simply found that motors
are unlikely to provide a mechanism for actin monomer transport. 

We hope that our results will spur further experimental investigations of 
cellular filopodia and their dynamics.


\section*{References}

\begin{thebibliography}{99}
\bibitem{bray} D. Bray, {\it Cell Movements: From Molecules to Motility}, 2nd
Edition, Garland Publishing New York (2001).
\bibitem{cellmotilityreview} D. A. Lauffenburger, A. F. Horwitz,
{\it Cell} {\bf 84}, 359 (1996).
\bibitem{peskin93} C. S. Peskin, G. M. Odell, G. F. Oster, 
{\it Biophys. J.} {\bf 65}, 316 (1993).
\bibitem{focaladhesions} K. Burridge, K. Fath, T. Kelly, G. Nuckolls,
C. Turner, {\it Ann. Rev. Cell Biol.} {\bf 4}, 487 (1988).
\bibitem{rhoda} R. J. Hawkins, R. Poincloux, O. Benichou,  
 M. Piel, P. Chavrier, R. Voituriez,
{\it Biophys. J.} {\bf 101}, 1041 (2011).
\bibitem{poincloux} R. Poincloux, O. Collin, F. Lizarraga, M. Romao, M. Debray, M. Piel, P. Chavrier,
{\it Proc. Natl. Acad. Sci. USA} {\bf 108}, 1943 (2011).
\bibitem{elsen}  E. Tjhung, D. Marenduzzo, M. E. Cates, 
{\it Proc. Natl. Acad. Sci. USA} {\bf 109}, 12381 (2012).
\bibitem{lamellipodium} 
T. M. Svitkina, G. G. Borisy, {\it J. Cell. Biol.} {\bf 145}, 1009 (1999).
\bibitem{svitkina} T. M. Svitkina, O. Y. Chaga, D. M. Vignjevic, S. Kojima, 
J. M. Vasiliev, G. G. Borisy, {\it J. Cell Biol.} {\bf 160}, 409 (2003).
\bibitem{mejillano} M.~R. Mejillano, S. Kojima,
D.~A. Applewhite, F.~B. Gertler, T.~M. Svitkina, G.~G. Borisy,
{\it Cell} {\bf 118}, 363 (2004).
\bibitem{mogilner} A. Mogilner, B. Rubinstein, {\it Biophys. J.}
{\bf 89}, 782-795 (2005).
\bibitem{daniels} D. R. Daniels, {\it Biophys. J.} {\bf 98}, 1139 (2010).
\bibitem{daniels2} D. R. Daniels, {\it Phys. Rev. Lett.} {\bf 100}, 048103 (2008).
\bibitem{matthew} D. R. Daniels, D. Marenduzzo, M. S. Turner, {\it Phys. Rev. Lett.} {\bf 97}, 098101 (2006).
\bibitem{papoian1} P. I. Zhuravlev, B. S. Der, G. A. Papoian, Biophys. J. 98, 1439-1448 (2010).
\bibitem{papoian2} Y. Lan, G. A. Papoian, {\it Biophys. J.} {\bf 94}, 
3839 (2008).
\bibitem{papoian3} P.~I. Zhuravlev, Y. Lan, M.~S. Minakova, G.~A. Papoian,
{\it Proc. Natl. Acad. Sci. USA} {\bf 109}, 10849 (2012).
\bibitem{papoian4} P.~I. Zhuravlev, G.~A. Papoian, {\it Cell. Adh. Migr.}
{\bf 5}, 448 (2011).
\bibitem{papoian5} P.~I. Zhuravlev, G.~A. Papoian, {\it Proc. Natl. Acad. Sci. USA} {\bf 106}, 11570 (2009).
\bibitem{actin} A. Ott, M. Magnasco, A. Simon, and A. Libchaber, 
{\it Phys. Rev. E} {\bf 48}, R1642 (1993).
\bibitem{eduardo} E. Sanz, D. Marenduzzo, {\it J. Chem. Phys.} {\bf 132},
194102 (2010).
\bibitem{ab04} S. S. Andrews, D. Bray, {\it Phys. Biol} {\bf 1}, 137 (2004).
\bibitem{ptvf92} 
W. H. Press, S. A. Teukolsky, W.T. Vetterling, B.P. Flannery, 
{\it Numerical Recipes in C: The Art of Scientific Computing}, Second
Edition, Cambridge University Press (1992).
\bibitem{carlsson01} A. E. Carlsson, {\it Biophys. J.} {\bf 81}, 
1907-1923 (2001).
\bibitem{sheetz92} M. P. Sheetz, D. B.Wayne, A. L. Pearlman,
{\it Cell Motil. Cytoskeleton} {\bf 22}, 160-169 (1992).
\bibitem{mcgrath98} 
J. L. McGrath, Y. Tardy, C.F. Dewey, J. J. Meister, J. H. Hartwig
{\it Biophys J.} {\bf 75}, 2070 (1998).
\bibitem{pollard86} T. D. Pollard, {\it J. Cell Biol.}, {\bf 103}, 
2747 (1986).
\bibitem{gardel08}
M. L. Gardel, B. Sabass, L. Ji, G. Danuser, U. S. Schwarz , C. M.
Waterman, {\it J. Cell. Biol.} {\bf 6}, 999 (2008).
\bibitem{elasticratchet1}
A. Mogilner, G. Oster, {\it Biophys. J.} {\bf 71}, 3030 (1996).
\bibitem{elasticratchet2}
N. J. Burroughs, D. Marenduzzo, {\it J. Chem. Phys.}  {\bf 123},
174908 (2005).
\bibitem{comet}
N. J. Burroughs, D. Marenduzzo, {\it Phys. Rev. Lett.}  {\bf 98},
 238302 (2007).
\bibitem{SEPR} 
K. E. P. Sugden, M. R. Evans, W. C. K. Poon, and N. D. Read {\it Phys. Rev. E} {\bf 75}
031909 (2007)
\bibitem{NFC}
S. A. Nowak, P.-W. Fok and T. Chou, {\it Phys. Rev. E} {\bf 76}
031135 (2007)
\bibitem{TEK}
J. Tailleur, M. R. Evans, and Y. Kafri, 
{\it Phys. Rev. Lett.} {\bf 102}, 118109 (2009). 
\bibitem{bc02} J. S. Berg, R. E. Cheney, {\it Nature Cell Biol.} 
{\bf 4}, 246 (2002).
\bibitem{bpc01} 
J. S. Berg, B. C. Powell, R. E. Cheney, {\it Mol. Biol. Cell} 
{\bf 12}, 780 (2001).
\bibitem{wwcmj96}
F. S. Wang, J. S. Wolenski, R. E. Cheney, M. S. Mooseker, D. G.
Jay, {\it Science} {\bf 273}, 660 (1996).
\bibitem{nrncbsr08} S. Nagy, B. L. Ricca, M. F. Norstrom, D. S. Courson, 
C. M. Brawley,
P. A. Smithback, R. S. Rock, {\it Proc. Nat. Acad. Sci. USA} {\bf 105}, 9616 
(2008).
\bibitem{e01} R. J. Ellis, {\it Trends Biochem. Sci.} {\bf 26}, 597 (2001).
\end{thebibliography}
\end{document}